\documentclass[final,3p,times,twocolumn]{elsarticle}
 \biboptions{comma,sort&compress}
\usepackage{here}
 \usepackage{graphicx}
  \usepackage{epsfig}

\def\nin{\noindent}
\def\beq{\begin{equation}}
\def\eeq{\end{equation}}
\def\bea{\begin{eqnarray}}
\def\eea{\end{eqnarray}}

\journal{Nuc. Phys. (Proc. Suppl.)}

\begin{document}

\begin{frontmatter}



\title{$W/Z$ + jet production at the Tevatron}

 \author[]{Lars Sonnenschein$^a$ \\
           \scalebox{0.9}{on behalf of the D\O\ and CDF collaborations}}
 \address[]{RWTH Aachen University \\
                   III. Phys. Inst. A \\
                   52056 Aachen, Germany.}



\begin{abstract}
\noindent
Vector boson plus jet production is interesting for Higgs search, 
beyond the Standard Model physics and provides standard candles for calibration.
This is complementary to inclusive jet production measurements which provide 
precision tests of perturbative QCD. A multitude of $W/Z$ plus heavy and 
light flavour jet measurements in $p\bar{p}$ collisions
at a centre of mass energy of $\sqrt{s}=1.96\,$TeV is discussed.
Next-to-Leading order perturbative QCD predictions and various models 
are compared to the measurements.
\end{abstract}




\end{frontmatter}


\section{Introduction}
\nin
The discussed measurements
are accomplished by the multi-purpose collider 
experiments D\O\ and CDF with their broad particle identification capabilities 
through central tracking, fine granulated calorimeters and muon spectrometers. 
The detectors are located at the Tevatron accelerator at Fermilab, where proton 
anti-proton collisions take place at a centre of mass energy of $1.96\,$TeV.
Vector boson plus jet production at the Tevatron is complementary to the 
kinematic regimes of the HERA accelerator and fixed target experiments.
It provides handles to Parton Distribution Functions (PDF), initial and 
final state gluon radiation and the transverse momentum ($p_T$) spectra of the 
jets and the vector bosons. The measurements provide standard candles for 
calibration and tuning of Monte-Carlo (MC) event generators. 
The understanding of these processes is important for 
Standard Model (SM) as well as Beyond the SM (BSM)
phenomenology. The Tevatron dataset is now large enough to confront predictions
and it has a unique kinematic overlap with the LHC and the expected SM Higgs mass 
range.  

The data of the measurements are fully corrected for instrumental effects. 
Therefore they can be directly used for testing and improving existing MC 
event generators and any future calculation and model.
Next-to-Leading Order (NLO) and Leading Order (LO) perturbative QCD (pQCD)
predictions from 
MCFM~\cite{mcfm} are compared to data taking non-perturbative
effects of hadronisation and underlying event from simulation into account
in the prediction, i.e. the comparison takes place at the hadronic final state
or synonymous at the particle level. The relative uncertainties of the 
measurements are dominated by the Jet Energy Scale (JES) uncertainty.
Iterative seed-based infrared safe midpoint cone jet algorithms are used by the 
D\O~\cite{d0cone} 
and CDF~\cite{cdfcone} 
experiments in Run~II.

In the following sections 2-6 the Tevatron vector boson plus light flavour jet 
measurements are presented, followed by the vector boson plus heavy flavour
jet measurements in sections 7-11.

\section{D\O\ inclusive $Z/\gamma^*(\rightarrow ee)$ + jet cross section}
\nin
The $Z$ + jet production cross section is measured~\cite{d0_plb678}
differentially in inclusive $p_T$ bins of leading, second and third 
leading jet, making use of an integrated luminosity of $1.0\,$fb$^{-1}$. 
Electron/positron candidates above a transverse momentum of
$p_T>25\,$GeV are selected in an absolute pseudorapidity range
of $|\eta_e|<1.1, 1.5<|\eta_e|<2.5$ to form $Z$ boson candidates in a mass window 
of $65<M_{ee}<115\,$GeV. Jets with a cone radius of $R=0.5$ and a 
$p_T^{\mbox{\scriptsize jet}}>20\,$GeV are selected in a rapidity range 
of $|y_{\mbox{\scriptsize jet}}|<2.5$.
The cross section is normalised to inclusive $Z$ + jet production which cancels 
uncertainties on luminosity and most of electron trigger and identification
uncertainties.
The phase space of the selected events is extrapolated to the full lepton 
kinematics.
While the (N)LO pQCD predictions are in agreement with data there are large 
differences between the different models of PYTHIA~\cite{pythia}, 
HERWIG~\cite{herwig} (+ JIMMY), ALPGEN~\cite{alpgen} + PYTHIA and 
SHERPA~\cite{sherpa}. The experimental errors are small and dominated by 
statistics, allowing for future improvements.

\section{CDF inclusive $Z/\gamma^*(\rightarrow ee)$ + jet cross section}
\nin
The $Z$ + jet production cross section is measured~\cite{cdf_prl100}
differentially in inclusive transverse momentum bins of the leading and second
leading jet, making use of an integrated luminosity of $1.7\,$fb$^{-1}$.
Electron/positron candidates above a transverse energy of $E_T^e>25\,$GeV are 
selected in an absolute pseudorapidity range
of $|\eta_e|<1.0, 1.2<|\eta_e|<2.8$ to form $Z$ boson candidates in a mass window 
of $66<M_{ee}<116\,$GeV. Jets with a cone radius of $R=0.7$ and a 
$p_T^{\mbox{\scriptsize jet}}>30\,$GeV are selected in a rapidity range 
of $|y_{\mbox{\scriptsize jet}}|<2.1$.
There is good agreement between data and NLO pQCD prediction.

\section{D\O\ inclusive $Z/\gamma^*(\rightarrow \mu\mu)$ + jet cross section}
\nin
Differential angular distributions of the $Z$ + jet production cross section is 
measured~\cite{d0_plb682} in inclusive bins of transverse $Z$ boson momentum
making use of an integrated luminosity of $1.0\,$fb$^{-1}$.
Oppositely charged muon candidates above a threshold of
$p_T>15\,$GeV are selected in an absolute rapidity range
of $|y_{\mu}|<1.7$ to form $Z$ boson candidates in a mass window 
of $65<M_{\mu\mu}<115\,$GeV. Jets with a cone radius of $R=0.5$ and a 
$p_T^{\mbox{\scriptsize jet}}>20\,$GeV are selected in a rapidity range 
of $|y_{\mbox{\scriptsize jet}}|<2.8$. The differential cross section 
in two inclusive bins of $Z$ boson transverse momentum ($p_T^Z>25\,$GeV and 
$p_T^Z>45\,$GeV)
is measured as a function of the angular variables rapidity sum, rapidity difference
and the azimuthal angle difference between the leading jet and the $Z$ boson.
ALPGEN~\cite{alpgen} and SHERPA~\cite{sherpa} include up to three partons 
in the matrix element calculations. The binning is chosen such, that the 
detector resolution causes little migrations between bins. There is less 
agreement between data and predictions in the lower 
$p_T^Z>25$ bin. This holds for the non-pQCD as well as for the pQCD predictions.
Among the non-pQCD predictions SHERPA (1.1.3) provides the best description
of the angular distributions, in particular $\Delta y(Z,j)$, in the
$p_T^Z>45\,$GeV bin. 
In the $p_T^Z>25\,$GeV bin the measured cross section is
$\sigma(Z+\mbox{jet})/\sigma(Z) =  \left[  47\pm 1(\mbox{stat.})\pm 2(\mbox{syst.}) \right]  \cdot 10^{-3}$ compared to the pQCD NLO prediction of
$\left[  40\pm 3(\mbox{scale})\pm 1(\mbox{PDF}) \right]  \cdot 10^{-3}$
and the pQCD LO prediction of
$\left[  40\pm 8(\mbox{scale})\pm 1(\mbox{PDF}) \right]  \cdot 10^{-3}$.

\section{CDF $Z(\rightarrow ee,\mu\mu)$ + 1 jet $p_T$ balance}
\nin
CDF measured the $Z$ boson + 1 jet exclusive $p_T$ balance~\cite{cdf_subNIM2010}, 
making use of an integrated luminosity of  $4.6\,$fb$^{-1}$
and sets precision limits on the measurements of Standard Model (SM) jet and
vector boson observables, relevant for the discovery potential of new physics. 
Oppositely charged lepton candidates ($\ell=e,\mu$) above a threshold of
$p_T^{\mu}, E_T^e>18\,$GeV are selected to form $Z$ boson candidates 
in a mass window of $80<M_{\mu\mu}<100\,$GeV. A leading jet 
with a $p_T^{\mbox{\scriptsize jet}}>8\,$GeV is selected in a pseudorapidity range 
of $0.2<|\eta_{\mbox{\scriptsize jet}}^{\mbox{\scriptsize det}}|<0.8$ to avoid cracks in 
the central calorimeter. The measurement is accomplished for three different
cone algorithm radii $R=0.4, 0.7, 1.0$. The azimuthal angle between the $Z$ 
boson and the jet has to satisfy $\Delta\phi>3.0\,$rad to provide back-to-back
configurations in the plane transverse to the beam axis. 
Predictions for $Z$ boson + jet production are obtained from
PYTHIA~\cite{pythia} and ALPGEN~\cite{alpgen}.
The CDF JES is calibrated via tuning of the calorimeter response to single 
particles. Therefore the $Z$ boson plus jet $p_T$ balance provides an independent 
test of the CDF JES.
\begin{figure}[tbh]
\centerline{
\includegraphics[width=7.5cm]{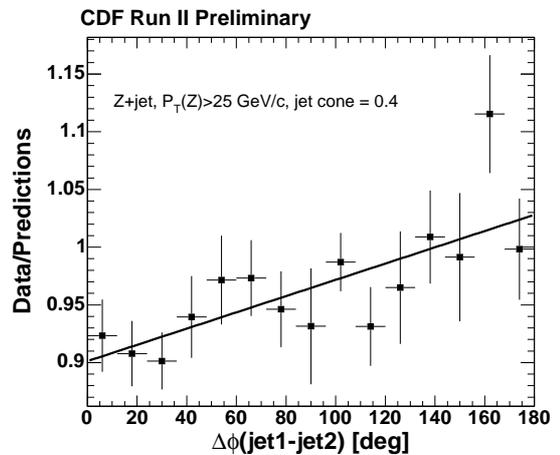}
}
\caption{ \label{deltaphiZjet}
Leading jet plus $Z$ boson $p_T$ balance data over simulation (PYTHIA) ratio 
as a function of the azimuthal angle difference between the leading and 
the second leading jet. Data has more sub-leading jets close in angle 
to the leading jet.
}
\end{figure}
While the measured track in jet distributions agree with the quark and gluon 
jet fractions of PYTHIA~\cite{pythia},
 a higher rate of sub-leading jets collinear to the 
leading jet is observed in data (see Fig. \ref{deltaphiZjet}). 
This does set limitations on the precision of
the JES determined at the ATLAS and CMS experiments by means of the jet plus $Z$ 
boson $p_T$ balance of the order of 3\%. Only large angle final state radiation
observed in form of sub-leading jets turns out to be able to explain the 
discrepancy.

\section{CDF inclusive $W(\rightarrow e\nu)$ + n=1-4 jets cross section}
\nin
The $W$ + $n$ jet inclusive production cross section ($n=1-4$) is 
measured~\cite{cdf_prd77}
differentially in inclusive transverse momentum bins of the $n$th leading jet, 
making use of an integrated luminosity of $320\,$pb$^{-1}$.
Electron/positron candidates above a transverse energy of $E_T^e>20\,$GeV are 
selected in an absolute pseudorapidity range
of $|\eta_e|<1.1$. A missing transverse energy of $E\!\!\!/\,_T>30\,$GeV and a 
transverse mass of $m_T^W>20\,$GeV is required to select $W$ boson candidates. 
Jets with a cone radius of $R=0.4$ and a 
$p_T^{\mbox{\scriptsize jet}}>20\,$GeV are selected in a pseudorapidity range 
of $|\eta_{\mbox{\scriptsize jet}}|<2.0$.
The measurement benefits of a production cross section which is about ten 
times higher than the one for $Z$ + jets production. At the same time the
multi-jet and top quark production background needs to be controlled.
There is good agreement between data and the NLO pQCD prediction. At low
transverse jet momentum the MC event generators ALPGEN~\cite{alpgen} plus
PYTHIA~\cite{pythia} (with MLM matching of the matrix element to the parton 
shower) and MADGRAPH~\cite{madgraph} plus HERWIG~\cite{herwig} (with CKKW 
matching) need a better modelling of the underlying event.

\section{D\O\ inclusive $\sigma(Z + b)/ \sigma(Z + j)$ ratio $(Z\rightarrow ee, \mu\mu)$}
\nin
The inclusive $Z$ + $b$ jet production cross section is measured~\cite{d0_6053conf}
as the ratio over the inclusive $Z$ + jet production cross section,
making use of an integrated luminosity of $4.2\,$fb$^{-1}$.
$Z$ boson decays into a pair of charge conjugated electrons or muons are considered.
Muon candidates with $p_T^{\mu}>10\,$GeV in the pseudorapidity range of 
$|\eta_{\mbox{\scriptsize det}}|<2.5$ and electron/positron candidates with 
$p_T^{e}>15\,$GeV in the pseudorapidity range of $|\eta_{\mbox{\scriptsize det}}|<2.0$
are selected together with jets ($R=0.5$). The leading jet has to have a 
transverse momentum of $p_T^{\mbox{\scriptsize jet}}>20\,$GeV in a pseudorapidity 
range of $|\eta_{\mbox{\scriptsize det}}|<1.1$. 
The $b$ flavour jets are separated from $c$ and light flavour 
jets by means of a neural network 
jet tagging algorithm based on the longer mean lifetime of heavy flavour hadrons
with respect to light ones. The flavour fractions are determined by a fit of
lifetime variable templates to the data. The templates are obtained from
ALPGEN~\cite{alpgen} plus PYTHIA~\cite{pythia} while the cross sections are
taken from NLO pQCD calculations.
The measured total cross section ratio is
$\sigma(Z\! + \! b) / \sigma(Z \! + \! j) = 0.0176 \pm 0.0024 (\mbox{stat.}) \pm 0.0023 (\mbox{syst.})$ consistent with the NLO pQCD prediction of
$0.0184 \pm 0.0022$.

\section{CDF inclusive $\sigma(Z + b)/ \sigma(Z)$ ratio $(Z\rightarrow ee, \mu\mu)$}
\nin
The inclusive $Z$ + $b$ jet production cross section is measured~\cite{cdf_prd79}
as the ratio over the inclusive $Z$ production cross section,
making use of an integrated luminosity of $2.0\,$fb$^{-1}$. 
$Z$ boson decays into a pair of charge conjugated electrons or muons are considered.
The cross section is measured differentially in bins of jet and heavy flavour 
jet multiplicity, as functions of leading $b$ jet transverse energy and pseudorapidity 
and as a function of the $Z$ boson transverse momentum.
$b$ jets are identified by means of a secondary vertex tagging algorithm.
The leading and second leading leptons ($e,\mu$) are required to satisfy
$p_T^{\mu_1}, E_T^{e_1}>18\,$GeV and $p_T^{\mu_2}, E_T^{e_2}>10\,$GeV.
The invariant dilepton mass has to lie in the interval $76<M_{\ell\ell}<106\,$GeV.
Jets with a cone radius of $R=0.7$ and a $E_T^{\mbox{\scriptsize jet}}>20\,$GeV are selected
in a pseudorapidity range of $|\eta_{\mbox{\scriptsize jet}}|<1.5$. 
The measured cross section ratio
 $\sigma(Z\!+\!b)/\sigma(Z)=\left[ 3.32\pm 0.53(\mbox{stat.})\pm 0.42(\mbox{syst.}) \right]\times 10^{-3}$ is in agreement with NLO pQCD predictions. Both data and
theory have large uncertainties. While data uncertainties are statistics dominated the
large uncertainty of the theory prediction arises from the missing NLO pQCD prediction
for $Z+b\bar{b}$ production which in turn causes a large scale dependence.

\section{D\O\ inclusive $\sigma(W + c)/ \sigma(W + j)$ ratio $(W\rightarrow \ell\nu)$}
\nin
The inclusive $W$ + $c$ jet production cross section is measured~\cite{d0_plb666}
as the ratio over the inclusive $W$ + jet production cross section,
making use of an integrated luminosity of $1.0\,$fb$^{-1}$.
The leptonic $W$ boson decay with a muon/electron in the final state is considered.
The cross section is measured differentially in bins of jet $p_T$.
Lepton candidates with $p_T^{\ell}>20\,$GeV and events with a missing transverse energy 
of $E\!\!\!/_T>20\,$GeV are selected. 
Jets with a cone radius of $R=0.5$ and $p_T^{\mbox{\scriptsize jet}}>20\,$GeV are selected 
in a pseudorapidity range of $|\eta_{\mbox{\scriptsize jet}}|<2.5$. 
$W$ boson + $c$ quark production is sensitive to the $s$ quark PDF content up to
scales of $Q^2=10^4\,$GeV$^2$. The charge sign of the lepton from the $W$ boson decay is
opposite (OS) to the one of the $c$ quark. This information is exploited in tagging an 
oppositely charged muon in the jet from the $c$ quark fragmentation. The same sign (SS)
subtracted background is small ($\sim1\%$). The multi-jet background is determined 
from data.  
The measured cross section is corrected for acceptances and efficiencies.
It amounts to
$\sigma(p\bar{p}\rightarrow W + c\!-\!\mbox{jet})/\sigma(p\bar{p}\rightarrow W+\mbox{jets}) = 0.074 \pm 0.019(\mbox{stat.})^{+0.012}_ {-0.014} (\mbox{syst.})$ and is
within the large errors in agreement with the prediction of $0.044\pm 0.003$ from 
ALPGEN~\cite{alpgen} plus PYTHIA~\cite{pythia}.

\section{CDF semi-incl. $W + c$ jet cross section ($W\rightarrow\ell\nu)$}
\nin
The $W$ + $c$ jet exclusive production cross section is measured~\cite{cdf_prl100b}
for selected events with one or two jets, 
making use of an integrated luminosity of $1.8\,$fb$^{-1}$.
Lepton candidates with $p_T^{\ell}>20\,$GeV in the pseudorapidity range
of $|\eta_{\ell}|<1.1$ and events with a missing transverse energy of 
$E\!\!\!/_T>25\,$GeV are selected. 
Jets with a cone radius of $R=0.4$ and $p_T^{\mbox{\scriptsize jet}}>20\,$GeV
are selected in a pseudorapidity range of  $|\eta_{\mbox{\scriptsize jet}}|<1.5$.
Charm flavour jets are tagged by means of a soft muon in the jet.
The OS-SS subtracted data distributions are in good agreement with the prediction
of ALPGEN~\cite{alpgen} with PYTHIA~\cite{pythia}.
The total measured cross section amounts to
$\sigma(W_{(\rightarrow\ell\nu)}+c)=9.8\pm 2.8(\mbox{stat.})^{+1.4}_{-1.6}(\mbox{syst.})\pm 0.6(\mbox{lum.})$~pb and is in good agreement with the NLO pQCD prediction of
 $\sigma(W_{(\rightarrow\ell\nu)}+c)=11.0^{+1.4}_{-3.0}$~pb for
$p_T^{\mbox{\scriptsize $c$-jet}}>20$~GeV and $|\eta_{\mbox{\scriptsize $c$-jet}}|<1.5$.

\section{CDF semi-incl. $W + b$ jet cross section ($W\rightarrow\ell\nu)$}
\nin
The $W$ + $b$ jet exclusive production cross section is measured~\cite{cdf_prl104}
for selected events with one or two jets, 
making use of an integrated luminosity of $1.9\,$fb$^{-1}$.
The leptonic $W$ boson decay with a muon/electron in the final state is considered.
Lepton candidates with $p_T^{\ell}>20\,$GeV and $|\eta_{\ell}|<1.1$
and events with a missing transverse energy of $E\!\!\!/_T>25\,$GeV are selected. 
Jets with a cone radius of $R=0.4$ and $p_T^{\mbox{\scriptsize jet}}>20\,$GeV
are selected in a pseudorapidity range of  $|\eta_{\mbox{\scriptsize jet}}|<2.0$.
Bottom flavour jets are tagged by means of a secondary vertex tagging algorithm.
A maximum likelihood fit to the vertex mass distribution of tagged jets is
applied to extract the flavour fractions of the selected data sample. The flavour
templates are obtained from ALPGEN~\cite{alpgen} with PYTHIA~\cite{pythia} and
MADEVENT~\cite{madevent} for single top quark production.
The measured total cross section is $\sigma(W_{(\rightarrow\ell\nu)}+b)=2.74\pm 0.27(\mbox{stat.})\pm 0.42(\mbox{syst.})$~pb. This has to be compared to the NLO pQCD prediction of
$\sigma(W_{(\rightarrow\ell\nu)}+b)=1.22\pm 0.14$~pb and the LO prediction of $\sigma(W_{(\rightarrow\ell\nu)}+b)=0.78$~pb from ALPGEN. The measured cross section exceeds the NLO prediction by about three $\sigma$ standard deviation.

\section{Conclusions}
\nin
Many measurements of vector boson plus light and heavy flavour jet production have been presented. Perturbative QCD predictions by means of MCFM~\cite{mcfm} are in good agreement with data in all inclusive vector boson plus jet production measurements.
The predictions of ALPGEN and MCFM are about three $\sigma$ below the
exclusive $W$ boson plus $b$ jet cross section measurement of CDF, where a reliable
secondary vertex tagger has been used.
The jet over $Z$ boson $p_T$ balance measurement of CDF shows up limitations for the
Jet Energy Scale precision at LHC experiments of the order of 3\%. The $p_T$
imbalance can be explained by large angle final state radiation in form of collinear sub-leading jets.
There is no perfect Monte-Carlo event generator. This holds for PYTHIA and 
HERWIG+JIMMY as well as for SHERPA and ALPGEN which are superior to the former 
parton shower Monte-Carlo event generators.
In general the data are corrected for detector effects, so that they can be compared to predictions at the hadronic final state. This means that the data can be re-used
for the tuning of Monte-Carlo event generators at any time in the future. 

\section*{Acknowledgements}
\nin
Many thanks to the staff members at Fermilab, collaborating institutions
and in particular the D\O\ and CDF collaborations. 













\end{document}